\documentstyle[11pt,paspconf,epsf]{article}

\begin{document}

\title{Photometric Redshifts in Hubble Deep Field South}
\author{D.L. Clements}
\affil{Dept. Physics and Astronomy, Cardiff university, PO Box 913, Cardiff,
CF22 3YB, UK}

\begin{abstract}
We apply both a traditional `dropout' approach and a photometric
redshift estimation technique to the Hubble Deep Field South data.  We
give a list of dropout selected z$\sim$3 objects, and show their
images. We then discuss our photometric redshift estimation technique,
demonstrate both its effectiveness and the role played by near-IR
data, and then apply it to HDF-S to obtain an estimated redshift
distribution.
\end{abstract}


\section{Introduction}

Techniques for photometric redshift estimation have advanced
considerably since the original Hubble Deep Field (North) (HDF-N)
observations (Williams et al., 1996). Initial techniques for
identifying candidate high redshift objects in HDF-N were inspired by
the famous `dropout' techniques originally developed by Steidel et
al. (see paper in these proceedings, and references therein).  These
have proved highly successful, and have identified several galaxies at
z$>$5 (eg. Weymann et al., 1998, Spinrad et al., 1998). Since then,
more sophisticated approaches to redshift estimation have been
developed (eg. Fernandez-Soto et al., 1998) which aim to determine a
redshift estimate rather than just select objects likely to have a
redshift in some broad range (eg. 2.5$<$z$<$3.5 for F300W dropouts in
one of the Hubble Deep Fields). The advent of the Hubble Deep Field
South (HDF-S) observations (Williams et al., 1999), which include not
only the optical WFPC-2 observations similar to HDF-N, but also STIS
UV/optical observations and NICMOS near-IR observations, allow us to
apply all these techniques to new fields and new wavelengths. We also
benefit from the extensive followup programme on HDF-N for validating
and training the various methods. The present paper describes the
application of both traditional `dropout' techniques, to select
candidate high redshift objects in HDF-S, and a `template matching'
photometric redshift estimation scheme for determining redshift
estimates for all the detected objects.

\section{Selecting High Redshift Candidates by the `Dropout' Method}

The simple `dropout' technique relies on the passage of the
912\AA~Lyman-limit discontinuity through broad-band imaging filters as
a function of redshift. At higher redshifts, the suppression of light
between 1216\AA~ and 912\AA~ in the emitted frame by the Ly$\alpha$
forest becomes significant, so that 1216\AA~ becomes the breakpoint in
the spectra (Spinrad et al., 1998). This allows one to select
candidate high redshift objects by their colours in broad band
filters. Typically, the objects are chosen by their absence in a blue
filter, and their relatively flat SEDs at redder wavelengths. The
classic application of this approach was by Steidel et al. (1995) who
used three filters, U, G and R, to select candidate z$\sim$3 objects
by their absence in the U filter. Similarly, 2.5$<$z$<$3.5 objects were
selected in HDF-N by their absence in the F300W filter (eg. Clements \& Couch,
1996). The technique can also be applied to longer wavelength filters
for the selection of still higher redshift objects. For example, the z=5.34
galaxy discussed in Spinrad et al. 1998 was discovered as a V-band
(F606W) dropout in the HDF-N images. 
The advent of HDF-S, which has coverage extending to 2.2$\mu$m in the
NICMOS field, thus offers the possibility of uncovering galaxies with
redshifts $>>$10 if we operate at the longest wavelengths.

Much has been learned about high redshift galaxies from experience
with HDF-N, so we can tune the selection criteria for HDF-S for
greater effectiveness. Consideration of the colour-colour plots in
Dickinson et al. (1998) suggests that the following criteria will
efficiently select candidate high redshift objects that drop out from
a given filter $a$ (all magnitudes referred to are in the AB system):
\begin{enumerate}
\item The colour between filter $a$ and the next reddest filter $b$ $>$1.5

\item The object should have been detected with at least 10$\sigma$
significance in filter $a$ if the colour $a - b = 1.5$

\item There is no more than 1 magnitude difference between the flux in
any adjacent filters redder than $b$

\end{enumerate}

However, at redder wavelengths, there are increasing possibilities for
foreground interlopers to turn up in these simple colour selections.
Specific problems include the 4000\AA--break in evolved galaxies, and
the unusual Very Red Objects. Both of these can masquerade as
`dropouts' especially at high redshift. Additional observations in the
near-IR for the WFPC-2 field, or optical for the NICMOS field would
help to reduce this contamination.

We apply these criteria to catalogs of HDF-S galaxies extracted from
the images using SExtractor (Bertin \& Arnouts, 1996). We uncover 15
candidate z$\sim$3 objects (F300W dropouts), 1 z$\sim$4 object (F450W
dropout), and 16 candidate z$\sim$5 (F606W dropouts) in the 4.7
sq. arcmin. WFPC-2 field. Further examination leads us to believe that
4 of the F300W objects are in fact nearby interlopers, and that there
is significant foreground contamination of the F606W dropout list. In
the NICMOS field we find 4 candidate z$\sim$6 (`optical' dropouts ---
absent from the STIS open-band optical image but present at F110W) one
candidate z$\sim$8 object (F110W dropout) and no candidate z$\sim$12
objects (F160W dropouts). One of the `optical' dropouts is the VRO
suspected to lie at z$>$1.7 discussed by Treu et al. (1998, 1999), and
we suspect other VROs contaminate the NICMOS list.

Images for the F300W dropouts are shown in Fig. 1 and basic parameters
given in Table 1. As can be seen, many of these objects appear to have
quite disturbed morphologies, similar to the F300W dropouts discovered
in HDF-N. Whether this is a true reflection of the underlying galaxy
morphology or a result of these observations lying in the rest-frame
far-UV is at this stage unclear. For more information on this work,
and for a full catalog of high redshift candidates, see Clements et
al. (1999).

\begin{table}
\begin{tiny}
\begin{tabular}{ccccccc} \hline
WFPC2 Cat. No.&RA(J2000)&Dec(J2000)&F814&F606&F450&F300\\ \hline
\multicolumn{7}{c}{z$\sim$ 3 F300 Dropout Candidates} \\ \hline
\\
  38&22 32 53.98&-60 34 20.53&24.76$\pm$0.08&24.96$\pm$0.04&25.08$\pm$0.13&26.60$\pm$0.6\\
 149$^1$&22 32 53.33&-60 34 13.69&22.83$\pm$0.02&23.21$\pm$0.01&23.88$\pm$0.04&26.14$\pm$0.26\\
 257&22 32 59.91&-60 34 05.02&24.14$\pm$0.02&24.90$\pm$0.02&25.29$\pm$0.06&27.01$\pm$0.47\\
 760$^1$&22 32 58.10&-60 33 37.08&23.13$\pm$0.01&23.60$\pm$0.01&24.29$\pm$0.03&26.28$\pm$0.29\\
 877&22 33 04.89&-60 33 29.48&24.31$\pm$0.03&24.43$\pm$0.01&24.68$\pm$0.04&$>$27.8\\
 880&22 32 50.64&-60 33 28.84&24.45$\pm$0.03&24.72$\pm$0.02&25.07$\pm$0.05&$>$27.8\\
 895&22 33 03.19&-60 33 28.77&23.34$\pm$0.02&23.68$\pm$0.01&24.02$\pm$0.03&26.20$\pm$0.25\\
1158&22 33 03.87&-60 33 12.89&24.56$\pm$0.04&24.77$\pm$0.02&25.15$\pm$0.07&$>$26.6\\
1562&22 32 57.22&-60 32 41.50&24.55$\pm$0.04&24.78$\pm$0.02&25.03$\pm$0.07&27.00$\pm$0.69\\
1745&22 32 49.00&-60 32 26.85&23.44$\pm$0.02&23.61$\pm$0.01&24.14$\pm$0.03&$>$27.8\\
1748&22 32 49.22&-60 32 27.09&23.00$\pm$0.01&23.20$\pm$0.01&23.68$\pm$0.03&$>$26.9\\
1816&22 32 47.62&-60 32 20.61&24.43$\pm$0.03&24.84$\pm$0.02&25.26$\pm$0.06&$>$26.9\\
1847$^1$&22 32 47.76&-60 32 17.59&22.72$\pm$0.01&23.15$\pm$0.01&23.72$\pm$0.03&25.71$\pm$0.16\\
1934$^1$&22 32 46.51&-60 32 08.38&22.97$\pm$0.02&23.40$\pm$0.01&24.07$\pm$0.04&26.61$\pm$0.41\\
1951&22 32 48.87&-60 32 06.71&24.11$\pm$0.02&24.32$\pm$0.01&24.54$\pm$0.04&26.07$\pm$0.21\\
\\ \hline
\end{tabular}
\end{tiny}
\caption{Details of drop out galaxies}
Limits are 3$\sigma$ upper limits. $^1$ indicates likely low redshift
interloper.
\end{table}

\begin{figure}
\plotone{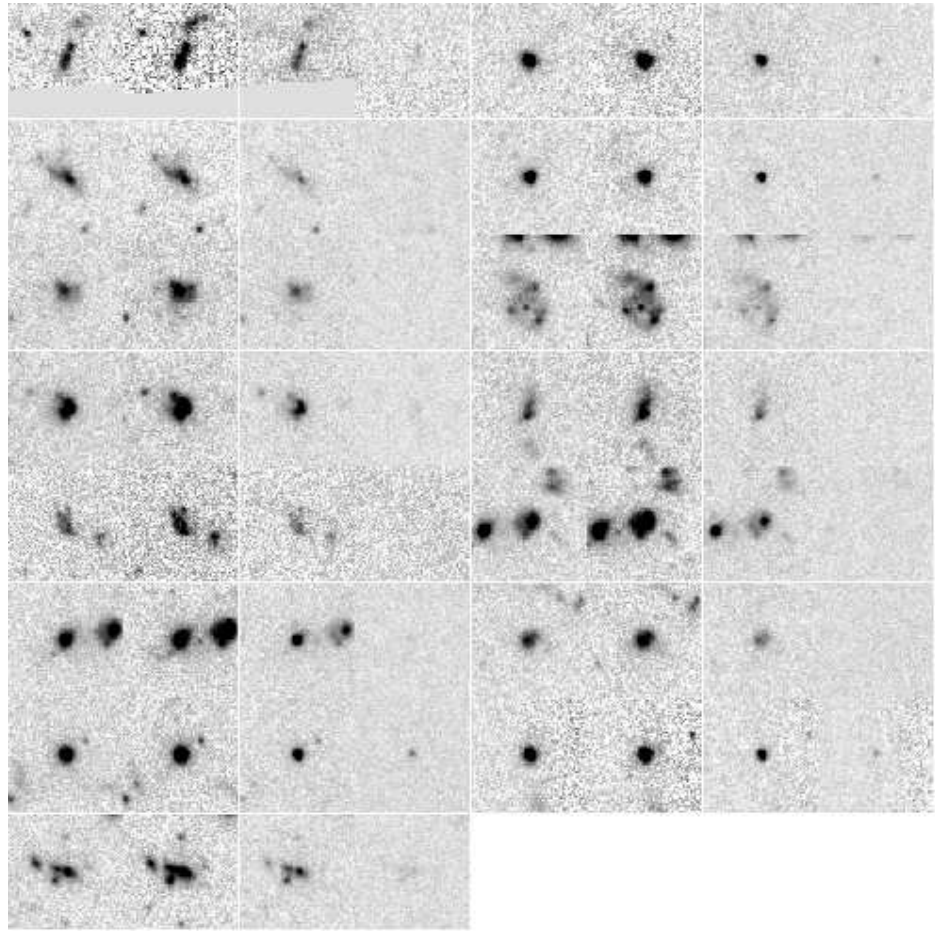}
\label{images}
\caption{Images of F300W dropouts. Images are shown in groups of four horizontally, showing
F814W, F606W, f450W and F300W images from left to right. Objects are shown in the same order
as in Table 1, starting in the top left, then left to right, and then working down the page.
Images are 4 arcseconds on a side. North is up and East is to the right.}
\end{figure}

\section{Photometric Redshift Techniques}

A more recent approach to multicolour surveys such as HDF-S is to
obtain a full photometric redshift solution. There are a number of
approaches to this, as can be seen from many of the other papers in
these proceedings. The approach adopted here, similar to that of
Fernandez-Soto et al. (1998), is to match a template spectral
energy distribution (SED) at a given redshift to the photometric data
for a galaxy. The likelihood that a galaxy is at the assumed redshift
can then be calculated:

\[ ln(L) = \sum_{filters} - \left ( \frac{F_{obs} - F_{mod}(z,n)}{E_{obs}} \right )^2
\]
where F$_{obs}$ is the observed flux in a given filter, F$_{mod}(z)$
is the flux predicted for this filter from the template SED given a
redshift $z$ and a normalisation $n$, and E$_{obs}$ is the error in the
observed flux. This calculation of likelihood makes the implicit
assumption that the flux errors are Gaussian.  This likelihood is then
maximised to find the most likely redshift for the galaxy given the
assumed template and certain astrophysical constraints for the
galaxy's properties. A selection of templates must be used. In the
present work we use six templates: elliptical, Sab, Scd and Irregular
types with SEDs taken from Coleman, Wu \& Weedman (1980) and
extrapolated to the infrared with GISSEL model SEDs (Bruzual \&
Charlot, 1996), and two starburst types with differing extinctions
taken from Kinney et al. (1996), again with IR extrapolation using
GISSEL models. We account for the effects of the Ly$\alpha$ forest
absorption using a parametrisation based on Madau (1995).

The effectiveness of this method can be tested using the HDF-N data
for objects which have spectroscopic redshifts. 108 such objects
appear in the compendium of Fernandez-Soto et al. (1998), and are used
as a test data set. The compendium includes infrared fluxes in J, H
and K for the objects and these are particularly useful for
constraining redshifts. Figure 2 shows the effectiveness of our
photometric redshift technique both with and without the near-IR
observations. A number of improvements to the method are still
possible, for example we are developing a better treatment for the
Lyman-forest absorption. We thus expect the estimations to become more
accurate. However, the method can clearly already be used as a fairly
reliable selector of candidate high redshift objects.

\begin{figure}
\plottwo{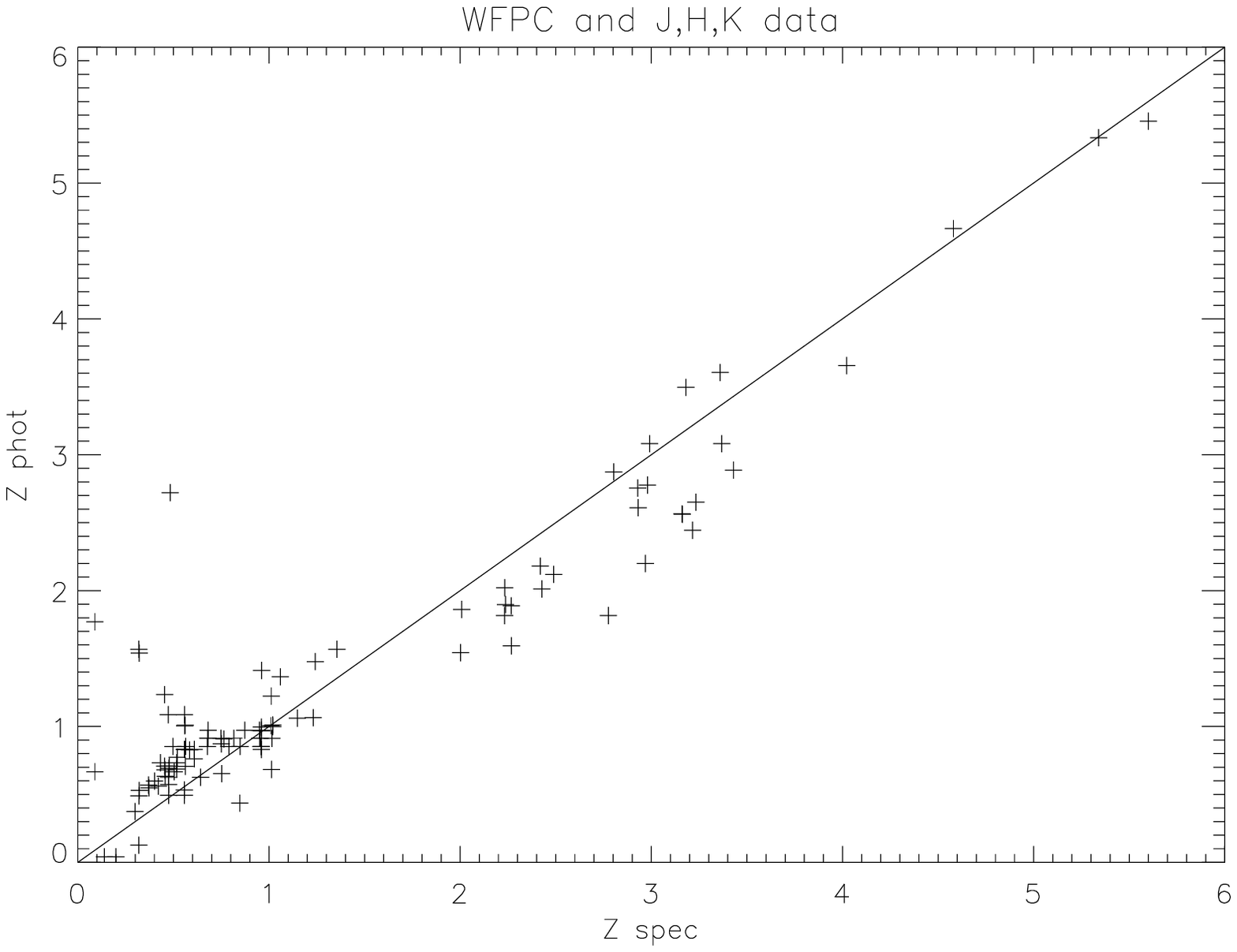}{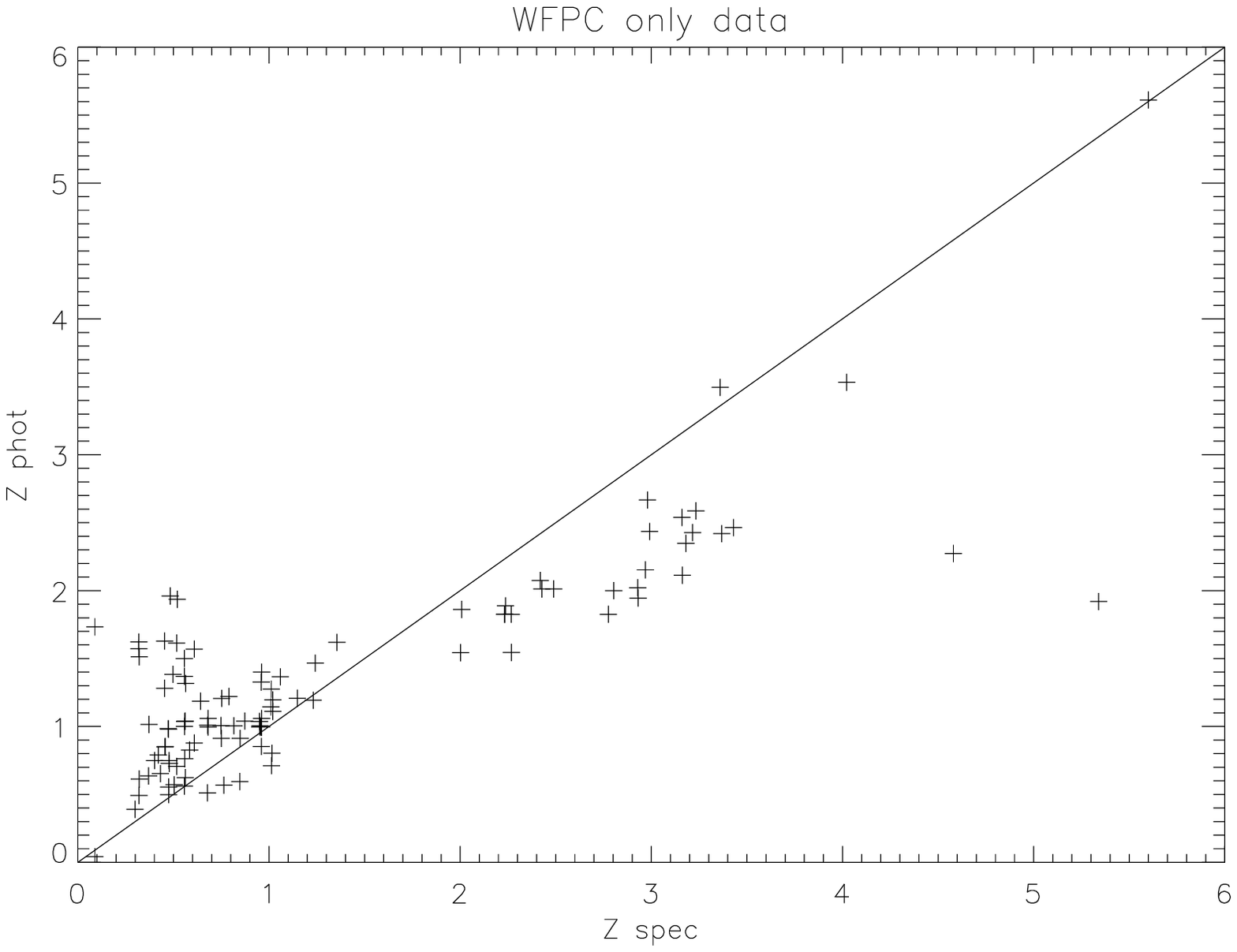}
\caption{Test of photometric redshift estimation}
Left, with near-IR data, right with only WFPC-2 data. 108 galaxies
from Fernandez-Soto et al. (1998) with spectroscopic redshifts and
fluxes at F300W, F450W, F606W, F814W, J, H and K are used for this test.
\end{figure}

We then apply our photometric redshift technique to the current
catalogue of objects in HDF-S (Williams et al., 1999). At present we
only include the WFPC-2 data, which will limit the accuracy of our
redshift estimates. IR will soon be added to the catalogue. In Figure
3 we present the redshifts distribution for HDF-S derived from our
photometric redshift estimation.

\begin{figure}
\plotone{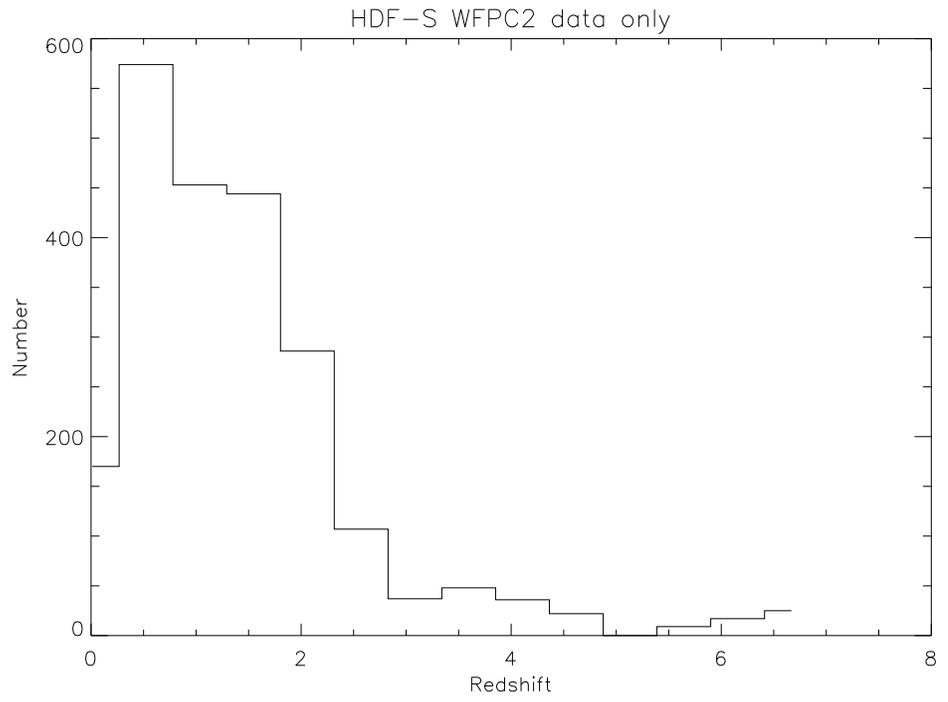}
\caption{Photometricly estimated redshift distribution for HDF-S}
\end{figure}

\acknowledgments

\end{document}